\documentstyle[12pt]{article}
\def\beqa{\begin{eqnarray}}
\def\eeqa{\end{eqnarray}}
\def\beq{\begin{equation}}
\def\eeq{\end{equation}}
\begin{document}
\begin{titlepage}
   \title{Quantum Violations of the Equivalence Principle
    in a Modified Schwarzschild Geometry. Neutrino Oscillations}
\author{V. Bozza$^{a, b}$\thanks{E-mail: valboz@sa.infn.it},
G. Lambiase$^{a, b}$\thanks{E-mail: lambiase@sa.infn.it}, G.
Papini$^{c, e}$\thanks{E-mail: papini@uregina.ca}, and G.
Scarpetta$^{b, d, e}$\thanks{E-mail: scarpetta@sa.infn.it} \\
 { \small{\em  $^a$  Dipartimento  di Scienze Fisiche ``E.R. Caianiello'',
   Universit\`a di Salerno, 84081 Baronissi (Sa), Italy.} }\\
 {\small {\em $^b$Istituto Nazionale di Fisica Nucleare, Sez. di
 Napoli, Italy.}} \\
 {\small{\em $^c$Department of Physics, University of Regina,
 Regina, Sask. S4S 0A2, Canada.}} \\
 {\small{\em $^d$Dipartimento di Fisica, Universit\'a di Salerno, 84081 Baronissi (Sa),
        Italy.}} \\
 {\small{\em $^e$International Institute for Advanced Scientific Studies,
 Vietri sul Mare (Sa), Italy.} }
 }
\date{\today}
\maketitle
\begin{abstract}
Neutrino flavour oscillations are analyzed in a model in which
particles experience an effective Schwarzschild geometry modified
by maximal acceleration corrections. These imply a quantum
violation of the equivalence principle. The corresponding shifts
in the phase of the neutrino mass eigenstates are calculated and
discussed.
\end{abstract}

\thispagestyle{empty}

\vspace{20.mm}


\vfill

\end{titlepage}

Einstein's equivalence principle plays a fundamental role in the
construction and testing of theories of gravity. Though verified
experimentally to better than a part in $10^{11}$ for bodies of
macroscopic dimensions \cite{roll,panov}, doubts have at times
been expressed as to its validity down to microscopic scales. It
is conceivable, for instance, that the equality of inertial and
gravitational mass break down for antimatter
\cite{goldman,goldman1}, or in quantum field theory at finite
temperatures \cite{donoghue}. Quantum violations of the
equivalence principle have also been discussed in Ref.
\cite{ahluwalia}

Einstein's equivalence principle is also violated in a model
developed by Caianiello and collaborators \cite{cai3}--\cite{cai5}
as a first step toward the unification of quantum mechanics and
general relativity. The model interprets quantization as curvature
of the eight-dimensional space-time tangent bundle TM. In this
space the standard operators of the Heisenberg algebra are
represented as covariant derivatives and the quantum commutation
relations are interpreted as components of the curvature tensor.

The merits of the model are extolled by its intrinsic simplicity
and the connections it establishes among seemingly unrelated
research areas. For instance, the model incorporates the notion
that the proper acceleration of massive particles has an upper
limit ${\cal A}_m$. Classical and quantum arguments supporting the
existence of a maximal acceleration (MA) have long been given
\cite{cai6}--\cite{pati}. MA also appears in the context of Weyl
space \cite{papini}--\cite{papiniM}  and of a geometrical analogue
of Vigier's stochastic theory \cite{vigier}.

${\cal A}_m$ is regarded by some as a universal constant fixed by
Planck's mass \cite{brandt}--\cite{gasp}, but a direct application
of Heisenberg's uncertainty relations \cite{caian,wood} and the
geometrical interpretation of the quantum commutation relations
given by Caianiello, suggest that ${\cal A}_m$ be fixed by the
rest mass of the particle itself according to ${\cal
A}_m=2mc^3/\hbar$. It is precisely through ${\cal A}_m$ that the
equivalence principle is violated.

MA delves into a number of questions. The existence of a MA would
rid black hole entropy of ultraviolet divergencies
\cite{hooft}--\cite{McG}, and circumvent inconsistencies
associated with the application of the point-like concept to
relativistic quantum particles \cite{lambiase,heg}.

A limit on the acceleration also occurs in string theory. Here the
upper limit manifests itself through Jeans-like instabilities
\cite{GSV,ven}  which occur when the acceleration induced by the
background gravitational field is larger than a critical value
$a_c = (m\alpha)^{-1}$for which the string extremities become
causally disconnected \cite{gasp1,gasp2}. $m$ is the string mass
and $\alpha$ is the string tension. Frolov and Sanchez
\cite{frolov} have then found that a universal critical
acceleration $a_c$ must be a general property of strings. It is
moreover possible to derive from it the generalized uncertainty
principle of string theory \cite{string}.

Applications of Caianiello's model include cosmology
\cite{cqg,gasp}, where the initial singularity can be avoided
while preserving inflation, the dynamics of accelerated strings
\cite{feoli}, the energy spectrum of a uniformly accelerated
particle  \cite{osc}, the periodic structure as a function of
momentum in neutrino oscillations \cite{osc} and the expansion of
the very early universe \cite{gasp1,gasp2}.

The extremely large value that ${\cal A}_m$ takes for all known
particles makes a direct test of the model difficult. Nonetheless
a direct test that uses photons in a cavity has also been
suggested \cite{pfs}. More recently, we have worked out the
consequences of the model for the classical electrodynamics of a
particle \cite{flps}, the mass of the Higgs boson
\cite{higgs,kuwata} and the Lamb shift in hydrogenic atoms
\cite{lamb}. In the last instance, the agreement between
experimental data and MA corrections is very good for $H$ and $D$.
For $He^+$ the agreement between theory and experiment is improved
by $50\%$ when MA corrections are included. MA effects in muonic
atoms appear to be measurable in planned experiments \cite{muon}.
MA also affects the helicity and chirality of particles
\cite{chen}.

Very recently we have studied the behaviour of classical
\cite{schw} and quantum \cite{qschw} particles in a Schwarzschild
field with MA modifications. In all these works space-time is
endowed with a causal structure in which the proper accelerations
of massive particles are limited. This is achieved by means of an
embedding procedure pioneered in \cite{osc} and further discussed
in \cite{schw}. The procedure stipulates that the line element
experienced by an accelerating particle is represented by
\begin{equation}\label{eq1}
 d\tau^2=\left(1+\frac{g_{\mu\nu}\ddot{x}^{\mu}\ddot{x}^{\nu}}{{\cal A}_m^2}
 \right)g_{\alpha\beta}dx^{\alpha}dx^{\beta}\equiv
 \sigma^2(x) g_{\alpha\beta}dx^{\alpha}dx^{\beta}\,{,}
 \end{equation}
and is therefore observer-dependent as conjectured by Gibbons and
Hawking \cite{hawking}. As a consequence, the effective space-time
geometry experienced by accelerated particles exhibits
mass-dependent corrections, which in general induce curvature, and
give rise to a mass-dependent violation of the equivalence
principle. The classical limit $\left ({\cal A}_m\right)^{-1} =
{\hbar\over 2 m c^3}\rightarrow 0$ returns space-time to its
ordinary geometry.

In Eq. (\ref{eq1}) $\ddot{x}^{\mu}=d^2x^{\mu}/ds^2$ is the, in
general, non--covariant acceleration of a particle along its
worldline. Caianiello's effective theory is therefore
intrinsically non-covariant.  Nonetheless the choice of
$\ddot{x}^{\mu}$ is supported by the derivation of ${\cal A}_m$
from quantum mechanics, by special relativity and by the weak
field approximation to general relativity.

The embedding procedure also requires that $\sigma^2(x)$ be
present in (\ref{eq1}) and that it be calculated in the same
coordinates of the unperturbed gravitational background. The model
is not intended, therefore, to supersede general relativity, but
rather to provide a way to calculate the quantum corrections to
the structure of space-time implied by Eq.(\ref{eq1}).

In this work we ask ourselves whether the violations of Einstein's
equivalence principle mentioned above have observable
consequences.

The best opportunity to observe an effect of this kind is perhaps
in connection with neutrino oscillations, as pointed out by
Gasperini \cite{gasp-osc} and Halprin and Leung \cite{halprin}.

Neutrino oscillations can occur in vacuum if the eigenvalues of
the mass matrix are not all degenerate and the corresponding mass
eigenstates differ from the weak eigenstates. Flavour oscillations
have frequently been advocated as possible explanations of the
solar neutrino deficiency and of the atmospheric neutrino problem.
The most discussed version of this type of solutions is the MSW
effect \cite{MSW,MSW1} in which the oscillations are enhanced by
matter in the Sun's interior.

Gravitational fields per se can not induce neutrino oscillations
because gravity couples universally to all kinds of matter.
Neutrino oscillations can however be induced by violations of the
equivalence principle. In this case oscillations arise if the
coupling of neutrinos to gravity is non-diagonal relative to
neutrino flavours. Only two--neutrino oscillations will be
considered in this work. A related calculation \cite{osc} was
performed for the two--dimensional problem of particles in
hyperbolic motion in a Kruskal plane. These particles are static
relative to Schwarzschild coordinates. This restriction is removed
below.

For convenience, the natural units $\hbar = c = G = 1$ are used
below. The conformal factor can be easily calculated as in
\cite{schw} starting from (2), with $\theta =\pi/2$, and from the
well known expressions for $\ddot{t},\ddot{r}$ and $\ddot{\phi}$
in Schwarzschild coordinates \cite{wh}. One obtains
 \[
\sigma^2(r)=1+\frac{1}{{\cal A}_m^2}\left\{
-\frac{1}{1-2M/r}\left(-\frac{3M\tilde{L}^2}{r^4}+\frac{\tilde{L}^2}{r^3}
-\frac{M}{r^2}\right)^2 + \right.
 \]
\begin{equation}\label{2}
\left. +\left(-\frac{4\tilde{L}^2}{r^4}+\frac{4\tilde{E}^2
M^2}{r^4(1-2M/r)^3}
\right)\left[\tilde{E}^2-\left(1-\frac{2M}{r}\right)\left(1+
\frac{\tilde{L}^2}{r^2}\right)\right]\right\}\,{,}
\end{equation}
where $M$ is the mass of the source, $\tilde{E}$ and $\tilde{L}$
are the total energy and angular momentum per unit of test
particle rest mass $m$. In the weak field approximation, the
modifications to the Schwarzschild geometry experienced by {\it
radially} accelerating neutrinos follows from (\ref{2}). One gets
\begin{equation}\label{2a}
\sigma^2(r)=1-\frac{1}{{\cal A}_m^2}\left(\frac{1}{4}+\frac{E^2}{m^2}
-\frac{E^4}{m^4}\right)\frac{r_s^2}{r^4}\,,
\end{equation}
where $r_s=2M$ is the Schwarzschild radius and $E$ is the total
energy. We neglect spin contributions. The effective Hamiltonian
for two--neutrino oscillations can be derived from the
Klein--Gordon equation of Ref. \cite{qschw}. Ignoring terms
proportional to the identity matrix and derivatives of $\sigma$,
one obtains
\begin{equation}\label{ham}
  H\sim \sqrt{E^2-m^2\sigma^2}\sim E-\frac{m^2\sigma^2}{2E}\,,
\end{equation}
in the approximation $m^2\sigma^2/2E^2<1$. The additional
corrections to the conventional two--neutrino oscillations are
therefore given by
\begin{equation}\label{corr}
  \frac{m^2}{2E{\cal
  A}^2_m}\left(\frac{1}{4}+\frac{E^2}{m^2}-\frac{E^4}{m^4}\right)\frac{r^2_s}{r^4}\,.
\end{equation}
The first term in (\ref{corr}) is independent of $m$ and will be
dropped. The second term is compatible with the approximation for
$r>(r_s^2/4m^2)^{1/4}\equiv r_{c1}$ which gives $r>2\times
10^{-2}$m if $r_s$ refers to the Sun and $m\sim 0.1$eV. If $r_s$
is that of Earth, the condition becomes $r>2\times 10^{-5}$m. The
last term in (\ref{corr}) satisfies the compatibility condition
for $r>(E^2r_s^2/4m^4)^{1/4}\equiv r_{c2}$ or $r>3\times 10^3$m
for $m\sim 0.1$eV, $r_s\sim 10^3$m (Sun) and $E\sim 1$GeV. Lower
values of $E$ and $r_s$ make the condition easier to meet. One
also finds $r_{c2}=(E/m)^{1/2}r_{c1}$.

Following Refs. \cite{gasp-osc,halprin} and taking into account
{\it medium effects}, one finds
\begin{equation}\label{evolution}
  i\frac{d}{dt}\left(\begin{array}{c}
                          \nu_e \\
                          \nu_\mu \end{array} \right)=
                          \frac{1}{2} \left(\begin{array}{cc}
                          2\sqrt{2}\,G_FN_e(r)-\displaystyle{\frac{\tilde{\Delta}}{r^4}}
                              \cos 2\theta & \qquad \displaystyle{\frac{\tilde{\Delta}}{r^4}}
                              \sin 2\theta \\
                          \displaystyle{\frac{\tilde{\Delta}}{r^4}}
                              \sin 2\theta & \qquad 0  \end{array} \right)\left(\begin{array}{c}
                          \nu_e \\
                          \nu_\mu \end{array} \right)\,,
\end{equation}
where
 \begin{equation}\label{6a}
 \tilde{\Delta}=\frac{\Delta
 m^2r^4}{2E}\left(1-\frac{r_s}{r}\right)+\frac{r_s^2E^3\Delta m^2(m_1^2+m_2^2)}{8m_1^4m_2^4}
 \left[1-\frac{m_1^2m_2^2}{E^2(m_1^2+m_2^2}\right]
 \end{equation}
and $\Delta m^2=m_2^2-m_1^2$. The second and forth terms in
(\ref{6a}) are just corrections of the other two and will be
dropped for simplicity. For ultra-relativistic neutrinos one has
$r\sim t$, and the equation of evolution can be re--cast in the
form
\begin{eqnarray}
 i\dot{\nu}_e &=& \left[\sqrt{2}\,G_FN_e(t)-\frac{\tilde{\Delta}
\cos 2\theta}{2 t^4}\right]\nu_e+\frac{\tilde{\Delta}}{2 t^4}\sin
2\theta \nu_\mu \label{ev1}\\
 i\dot{\nu}_\mu &=&\frac{\tilde{\Delta}\sin 2\theta}{2 t^4}\,
 \nu_e\,. \label{ev2}
\end{eqnarray}
Eqs. (\ref{ev1}) and (\ref{ev2}) can be de--coupled and the
equation of evolution of the flavour eigenstate $\nu_\mu$ is
 \begin{equation}\label{dis}
 \ddot{\nu}_\mu+\left[\frac{4}{t}+\frac{i}{2}\left(2\sqrt{2}G_FN_e(t)-
 \frac{\tilde{\Delta}\cos 2\theta}{t^4}\right)\right]\dot{\nu}_\mu+\left(\frac{\tilde{\Delta}\sin2
\theta}{2t^4}\right)^2\nu_\mu=0\,.
 \end{equation}
>From Eq. (\ref{evolution}) one derives the resonance condition
\begin{equation}\label{risonanza}
 \cos 2\theta =
 \frac{2\sqrt{2}\, G_F N_e(r) r^4}{\tilde{\Delta}}\,,
\end{equation}
where, for the Sun, $N_e(r)\sim N_0\exp (-10.54
r/R_{\odot})$cm$^{-3}$, $N_0=85 N_A$cm$^{-3}$. $N_A$ is Avogadro's
number. In the case of the Sun $\sqrt{2}G_fN_e\sim 10^{-12}$eV,
while for a supernova $\sqrt{2}G_FN_e\sim 1$eV. Eq.
(\ref{risonanza}) is therefore satisfied by $\cos 2\theta \approx
0$ even for high values of $r$. At resonance, Eq. (\ref{dis})
reduces to the form
\begin{equation}\label{final}
  \ddot{\nu}_\mu+\frac{4}{t}\dot{\nu}_{\mu}+\left(\frac{\tilde{\Delta}}{2t^4}
  \right)^2\nu_\mu=0\,.
\end{equation}
Taking $m_2\sim m_1\sim m$, one finds
 \begin{equation}\label{12bis}
 \tilde{\Delta}\approx \frac{\Delta m^2 r^4}{2E}+\frac{E^3r_s^2}{4}\,\frac{\Delta
 m^2}{m^6}\equiv \Delta \Phi_{(0)}+\Delta \Phi_{{\cal A}_m}\,.
 \end{equation}
The drastically different behaviours in $E$ and $r$ of the two
terms in (\ref{12bis}) now require some discussion. The phase
generated by the MA corrections is, in particular, proportional
to $r^{-3}$ which indicates its potential relevance at short
distances from the neutrinos source. The two terms become
comparable in size at a distance
$r_0=E(r_s/\sqrt{2}m^3)^{1/2}=(E\sqrt{2}/m)^{1/2}r_{c2}$. For
$r_{c2}<r<r_0$ the MA correction term predominates, but subsides
rapidly in the region $r>r_0>r_{c2}$ where the conventional term
takes over. There are therefore two possibilities to consider.
 \begin{description}
 \item[$i)$] $r_{c2}<r<r_0$. In this case the solution is
 \begin{equation}\label{sol}
  \nu_\mu=ae^{i\tilde{\Delta}/6t^3}+
 b e^{-i\tilde{\Delta}/6t^3}\,,
 \end{equation}
where $a$ and $b$ are constants and $\tilde{\Delta}\sim \Delta
\Phi_{{\cal A}_m}$. The oscillatory behaviour predominates when
$\tilde{\Delta}/12r^3\sim 1$, which gives the characteristic
length $L_{0}\sim (\tilde{\Delta}/12)^{1/3}$. For $r>L_0$, the
oscillations decrease rapidly. There is therefore a sphere of
radius
 \[
 L_0\sim \frac{E}{2m^2}\left(\frac{\Delta m^2
 r_s^2}{6}\right)^{1/3}>r_s
 \]
within which neutrino oscillations take place at a significant
rate. This sphere is external to the impenetrable shell discussed
in \cite{schw} and \cite{qschw}. The ratio
 \[
 \frac{L_0}{r_0}=\frac{(\Delta m^2)^{1/3}r_s^{1/6}}{3.1 m^{1/2}}
 \]
 becomes unity for $\Delta m^2\sim 28.5 \sqrt{m^3/r_s}$. If
 $\Delta m^2>28.5 \sqrt{m^3/r_s}$, then $r_0<L_0$ and this
 oscillation mechanism becomes less significant. This may be
 illustrated numerically as follows. In the case of  the Sun,
 $r_s\sim 10^3$m and for $m\sim 0.1$eV,
 $\Delta m^2\sim 10^{-2}$eV$^2$, $E\sim 10$MeV, one finds
 $r_0 \sim 6.2\times 10^5$m and $L_0\sim
 6.5 \times 10^7$m, which indicates that a negligible fraction
 of $\nu_e$ would be converted into $\nu_\mu$ by the mechanism discussed.

 For atmospheric neutrinos in the gravitational field of Earth
 ($r_s\sim 8\times 10^{-3}$m) and the values $E\sim 1$GeV, $m\sim
 0.1$eV, $\Delta m^2\sim 10^{-2}$eV$^2$ one obtains $r_0\sim L_0\sim 10^6$m
 which indicates appreciable conversion in regions of space
 surrounding Earth.

 \item[$ii)$] $r>r_0>r_{c2}$. For these values of $r$ the MA
 corrections become negligible, $\tilde{\Delta}\sim \Delta
 \Phi_{(0)}$ and Eq. (\ref{final}) becomes
 \begin{equation}\label{new}
 \ddot{\nu}_{\mu}+\frac{4}{t}\dot{\nu}_{\mu}+\omega^2\nu_\mu=0\,,
 \end{equation}
 where $\omega=\Delta m^2/4E$. The solution of (\ref{new}) is
 \begin{equation}\label{new1}
 \nu_\mu=\frac{f}{(\omega t)^{3/2}}Z_{-3/2}(\omega t)=
 \frac{f}{(\omega t)^2}\left(\sin \omega t +\frac{1}{\omega t}\,
 \cos \omega t\right)\,,
 \end{equation}
 where $f$ is a constant. The amplitude of the oscillations is
 damped for $r>4E/\Delta m^2$.
 \end{description}
In summary, the quantum violations of the equivalence principle
predicted by Caianiello's model lead to neutrino oscillations that
are characterized, at resonance, by two lengths, $r_0$ and $L_0$.
For $r<r_0$ the oscillations induced by MA dominate. $L_0$ is the
value of $r$ for which the induced phase gives a relevant
contribution. Ideally, one would have $r_0\sim L_0$. This
condition is satisfied if $\Delta m^2\simeq 28.5\sqrt{m^3/r_s}$
which favours larger neutrino masses and situations in which the
gravitational source has a small $r_s$. It is the case, for
instance, of atmospheric neutrinos in the gravitational field of
the Earth.

For $r>r_0$ the importance of the equivalence principle violations
induced by MA decreases rapidly and the conventional (damped)
vacuum oscillations dominate. Finally, the present calculations
underscore the basic point of Gasperini, Halprin and Leung that
violations of the equivalence principle are important in neutrino
oscillations. At the same time, the calculations also show that
this statement is not meant to apply universally and that a
detailed model of the violations may provide different results
for different gravitational sources, physical situations, regions
of space and values of the parameters involved.

 \vspace{0.5cm}

 \centerline{\bf Acknowledgements}

Research supported by MURTS fund PRIN 99 and the Natural Sciences
and Engineering Research Council of Canada. G.L. aknowledges the
financial support of UE (P.O.M. 1994/1999).

\end{document}